\documentclass[global,twocolumn]{svjour}
\usepackage{graphicx}
\usepackage{dcolumn}
\usepackage{bm}

\begin{document}


\title{A magnetic lens for cold atoms controlled by a rf field}

\author{ E. Mar\'echal, B. Laburthe-Tolra, L. Vernac, J.-C. Keller, and O. Gorceix}
\institute{Laboratoire de Physique des Lasers, UMR 7538 CNRS,
Universit\'e Paris Nord, 99 Avenue J.-B. Cl\'ement, 93430
Villetaneuse, France}

\maketitle

 \email{marechal@galilee.univ-paris13.fr}

\date{\today}
\begin{abstract}
We report on a new type of magnetic lens that focuses atomic clouds
using a static inhomogeneous magnetic field in combination with a
radio-frequency field (rf). The experimental study is performed with
a cloud of cold cesium atoms. The rf field adiabatically deforms the
magnetic potential of a coil and therefore changes its focusing
properties. The focal length can be tuned precisely  by changing the
rf frequency value. Depending on the rf antenna position relative to
the DC magnetic profile, the focal length of the atomic lens can be
either decreased or increased by the rf field.
\end{abstract}



\section{\label{sec:level1}Introduction}

In recent years, the combination of static inhomogeneous magnetic
fields with a rf field has been widely studied, theoretically and
experimentally in the quest of the realization of new trapping
geometries. As pointed out by Zobay and Garraway
\cite{zobay1,zobay2}, the use of a strong rf field allows to distort
strongly the static magnetic potentials and to create new adiabatic
potentials, with a much higher variety than standard magnetic traps.
Depending on the experimental arrangements and on the rf field
polarization, new traps have been proposed and some have been
demonstrated : bubbles, rings, double wells, lattices.
\cite{Colombe04,lesanovsky2006,fernholz2007,Courteille06}. All these
studies are based on an adiabatic deformation of a conventional
Ioffe Trap used to store a cold atom cloud or a Bose-Einstein
condensate. Here, we investigate an other issue, where a magnetic
atomic lens is tuned by a rf magnetic field. We have experimentally
investigated how the focal length of the lens can be controlled by
changing the frequency of the rf field. We show that using a rf
field in combination with conventional magnetic atom optics elements
\cite{Hinds99} adds flexibility to these components. The experiment
is done using cm range current carrying coils but could be
integrated in an atom chip component.

\section{\label{principles}Principle of the lens 'dressed' by rf}

The principle of the experiment is as follows. We use a spin
polarized cloud of cold cesium atoms. After a free fall time of
about 400 ms, the cloud enters the magnetic lens region. The rf
dressed lens is realized with two components: a DC field magnetic
lens, made of a simple coil, and a rf field. The inhomogeneous
static field of the magnetic lens defines a surface where atoms are
resonant with the rf field. As the atoms cross this interaction
surface, their spin orientation is changed. A spin flip occurs with
a probability close to one if the rf power is sufficient. Depending
on the initial polarization, the effect of the lens (initially
converging or diverging before the resonance surface) is reversed.
The magnetic lens is therefore separated by the rf interaction
surface in two parts, and becomes equivalent to a doublet. The
position of the interaction region, and therefore the focal of the
doublet, can be tuned by changing the rf frequency. The combination
of the permanent magnetic field with a rf field allows thus to
realize a tunable atomic lens. Note that contrary to our previous
works \cite{marechal99,miossec2002} the lens does not work in the
pulsed regime but is permanently fed with a constant current.

\begin{figure}[h]
\centering
\includegraphics[width=3in]{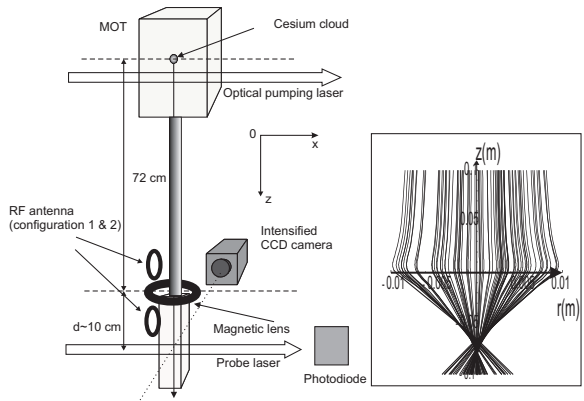}
\caption{\setlength{\baselineskip}{6pt} {\protect\scriptsize
Schematic of the experimental set-up. The lens is made of a vertical
axis coil, centered about 72 cm below the MOT center. The rf antenna
is located a few cm above (configuration 1) or below (configuration
2) the lens center, with its axis along the Ox direction. The inset
on the right shows 100 atomic trajectories as the cloud crosses the
magnetic lens plane.}} \label{ExperimentalSetUp}
\end{figure}

\section{\label{experimentalsetupsection}Experimental set-up}

The experimental set-up is presented on figure
\ref{ExperimentalSetUp}. We prepare a cloud of cold cesium atoms in
a standard magneto-optical trap. After a loading time of 1 s and a
molasses phase of 15 ms, we obtain a cloud of $10^7$ atoms, at a
temperature of 6 $\mu K$ with a 0.8  mm rms radius. The cloud is
then released and falls along the vertical axis. After a  2 cm free
fall, the cloud crosses a weak circularly polarized retroreflected
laser resonant with the $ |^7S_{1/2};F=4\rangle \rightarrow |^7
P_{1/2};F'=5\rangle $ transition. The atoms are then optically
pumped in the $ |F=4, m_F=+4\rangle$ or in the$ |F=4, m_F=-4\rangle$
Zeeman substate depending on the sign of the circular polarization.
During this polarization phase, a 5 G magnetic field is applied
along the laser axis. We obtain a polarization of $ 70\% $ of the
atoms in the aimed state, measured through an independent
longitudinal Stern Gerlach analysis \cite{Guibal99}. The cloud then
reaches the magnetic lens region, located  about $ 70$ cm below the
MOT center, and is focused. At the lens altitude, the cloud mean
velocity is $3.7$ m/s and it constitutes a pulsed atomic beam with a
narrow velocity spread $\Delta v/v \simeq 2\%$. Chromatic
aberrations are thus negligible in our experiment.

The magnetic lens is made of a 66 turn square coil wrapped on a
water cooled copper support. We have measured the magnetic field
along the $Oz$ axis and we have checked that close to the axis the
magnetic field created by the coil is equivalent to the one created
by a thin circular coil with 66 turns and with a 3.15 cm radius. For
the numerical simulations, we have replaced the square coil by its
equivalent circular coil with the previous parameters. The current
can be switched ON/OFF with a high power MOSFET. Our power supply
allows us to reach a maximum current of 100 A, corresponding to a
magnetic field value of 0.13 T at the coil center.

The rf antenna consists of a single turn coil of 2 $cm $ diameter,
with its axis along $Ox$. The coil is in series with a 50 $\Omega$
resistive load. This resistance dominates the total impedance of the
circuit and allows the matching between the antenna impedance and
the 10 $W$ rf amplifier output impedance. As explained later, the rf
antenna can be located above or below the lens center, leading to
two experimental configurations of the magnetic lens dressed by the
rf field. Furthermore the rf antenna diameter is much smaller than
the magnetic lens diameter, so that the rf interaction zone is
localized in a region close to the antenna position, where the rf
field is strong enough to induce adiabatic transitions.

\begin{figure}[h]
\centering
\includegraphics[width=3in]{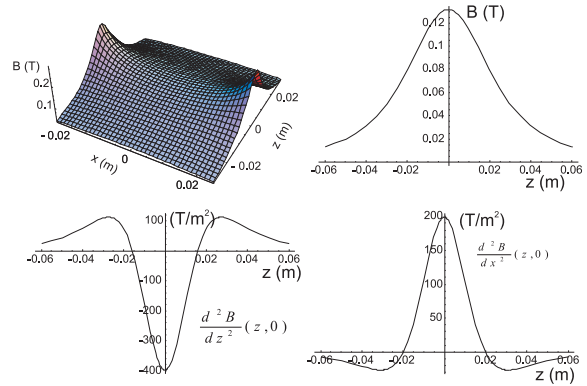}
\caption{\setlength{\baselineskip}{6pt} {\protect\scriptsize Space
dependence of the magnetic field amplitude and its second
derivatives. The magnetic field amplitude is plotted for a current
in the coil of 100 A.}} \label{ChampsMagnetique}
\end{figure}

The detection is made by imaging on an intensified CCD camera the
atom fluorescence due to the excitation by a resonant laser located
at a distance $d$ below the lens center ($d$ is about $ 10$ cm). The
magnetic field has to be switched off during the imaging time. Due
to the eddy current decay time, we have to wait for 5 ms more before
taking a picture. The probe is a horizontal retroreflected laser
light sheet along Ox in Lin $ \bot$ Lin configuration. It has a 1 mm
thickness and  a 10 mm width along Oy. At the same time, the probe
absorption is monitored on a photodiode, allowing us to measure the
cloud temperature by time of flight when the lens is switched off :
the cloud temperature of 6 $\mu K $ corresponds to a time of flight
width $(1/e^{2})$ of 7.9 ms. The image is taken by integrating the
fluorescence with the camera during 20 ms. This duration is long
enough to integrate all the atomic signal, even in presence of the
longitudinal diverging effect of the lens which increases the time
of flight width.

\section{\label{magneticlens}The magnetic lens}

Before describing the lens effect in combination with a rf field, we
first present the lensing effect without rf. For atoms in $ |F=4,
m_F=+4 \rangle$ the magnetic potential is $ \mu _B |B| $. The main
properties of this saddle like magnetic potential are summed up on
figure \ref{ChampsMagnetique}. As the magnetic field amplitude along
B is maximum at the coil center, atoms are accelerated then
decelerated as they cross the lens. The lens effect is caused by the
parabolic part of the magnetic potential, which is proportional to
the potential second derivatives. Along the vertical 0z axis, the
lens acts as a diverging lens, as the potential second derivative is
mainly negative. We will not investigate further this effect
\cite{marechal99}. Along the radial directions, the second
derivative is mainly positive and the lens acts as a converging
lens. We show on figure \ref{ExperimentalSetUp} (inset) the results
of a numerical simulation of the atoms trajectories as they cross
the magnetic lens. Atoms can also be initially prepared in the $
|m_F=-4 \rangle$ substate and the lens becomes diverging. In the
next section, we will use a rf field to resonantly spin flip the
atoms while they travel through the lens, and realize a lens with a
rf-controlled focal length.

\begin{figure}[h]
\centering
\includegraphics[width=3 in]{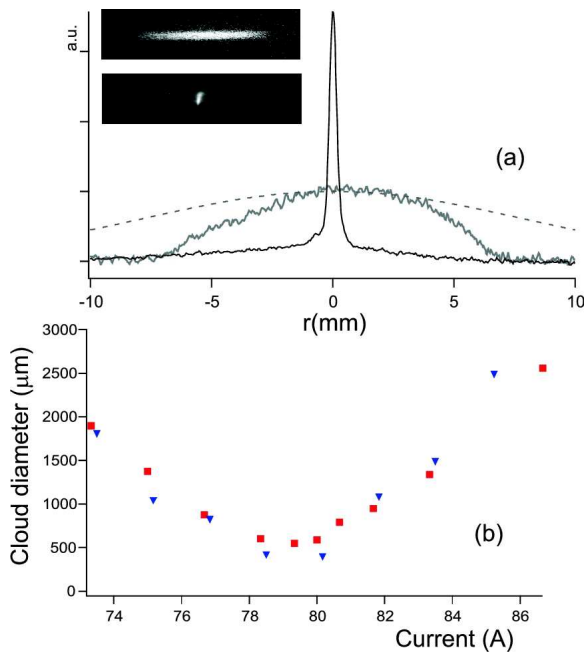}
\caption{\setlength{\baselineskip}{6pt} {\protect\scriptsize (a)
Fluorescence images of the cloud with the lens off and on, and
corresponding profiles along $Ox$ (grey and black). The current in
the magnetic lens is 79 $A$. The dashed line shows the expected
cloud profile in absence of truncation by the vacuum tubing. (b)
Cloud diameter at the probe position (7.5 $cm$ below the lens) as
function of the current in the lens. Experiment (squares) and
numerical simulation (triangles).}} \label{Focalization}
\end{figure}

On figure \ref{Focalization}, we show a fluorescence picture when
the cloud is focused at the probe position, located $ 7.5$ cm below
the coil center. We show also the fluorescence image of the cloud
when the lens is switched off. A gaussian fit of the profile gives a
500 $\mu m$ $1/e^2$ diameter of the focused cloud to be compared to
the 3.1 cm diameter of the free expanding cloud (80 $cm$ of free
fall at a temperature of 6 $\mu K$). The observed image without lens
is not gaussian and has a measured width of only $1.3$ $cm$. The
reason for this smaller than expected size is that the cloud is
truncated by an aperture of $0.8$ $cm$ diameter located half way
between the MOT cell and the lower glass cell; this aperture is used
as a differential pumping section. Futhermore, the cloud is also
truncated by the lower cell aperture of $1.6$ $cm$ diameter. As
shown on figure \ref{Focalization}(a), for a current of $79$ $A$,
the cloud is focused at the probe altitude. We have measured the
transverse diameter of the cloud  at the probe position as a
function of the current in the lens. The results are shown on figure
\ref{Focalization}(b). The cloud diameter is minimal as it is
focused at the probe position. For a higher current value, the cloud
is focused above the laser probe, for a lower value it is focused
below the laser beam. We have compared these experimental study with
a Monte Carlo numerical simulation based on the integration of 1000
classical trajectories of atoms in the magnetic potential. The 3D
initial positions and velocity of the atoms are chosen at random,
with the experimental gaussian distribution parameters. The
numerical results reproduce well the experimental outcomes. Note
that the minimal diameter value $D_{min}$ of the focused cloud,
about 500 $\mu m$, is limited by the magnetic lens spherical
aberrations; the initial cloud size contribution to $D_{min}$ being
negligible here.

\section{\label{DressedLens1}Magnetic lens dressed by rf}

\subsection{\label{DressedLens2}Principle}

In combination with a rf field, the lens properties are strongly
modified. We apply a strong monochromatic, linearly polarized  rf
field $ \overrightarrow{B} _{rf} \cos (\omega _{rf} t)$, which
couples the different sublevels $ |F=4, m_F\rangle \leftrightarrow
|F=4, m_F \pm 1\rangle$. The coupling creates an avoided crossing
between the different sublevels, and modifies the magnetic
potentials. The coupling strength is on the order of $ \Omega = g_F
\mu _B B_{rf}/2\hbar$ where $B_{rf}$ is the magnetic field
amplitude. The exact coupling strength depends on the rf amplitude,
on the relative orientation of the rf field with respect to the
static field, and is $m_F$ dependent. In our case, the coupling
dependence on the field orientation is optimal, the rf coupling
field $ \overrightarrow{B} _{rf}$ being perpendicular to the
vertical static magnetic field. As the static magnetic field is
inhomogeneous, the coupling is effective at positions where the
resonance condition is fulfilled that is when $ \hbar \omega
_{rf}=g_F \mu _B |B|$. This condition defines an isoB surface, an
atom crossing this surface is spin flipped. The correct frame to
treat the problem of the coupling of the atoms with the rf field is
the dressed atom picture. In presence of a rf field the nine
magnetic bare states $|F=4,m_F \rangle $ having initial potential
energies $E_{m_F}=m_Fg_F\mu_B B$ give rise to nine dressed states in
presence of rf with energies \cite{Colombe04}:
\begin{equation}\label{equ1}
    V_{m_F}=m_F\sqrt{(g_F\mu_B B-\hbar\omega_{rf})^2+(\hbar\Omega)^2}
\end{equation}

\begin{figure}[h]
\centering
\includegraphics[width=3.2 in]{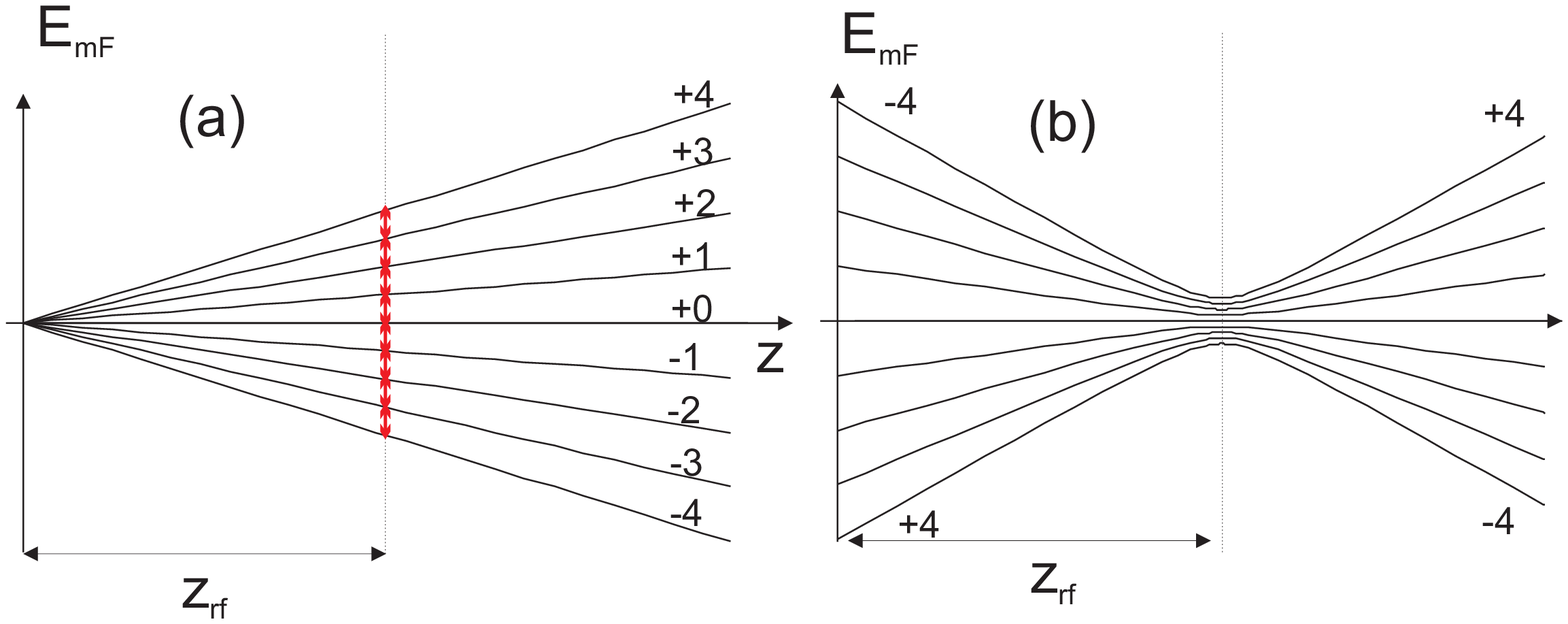}
\caption{\setlength{\baselineskip}{6pt} {\protect\scriptsize
Modification of the magnetic potential in presence of a strong rf
field with frequency $\omega_{rf}/2\pi $. (a) Initial potentials
with $B(0,0,z)=b'.z $. (b) The magnetic potentials are distorted
into nine adiabatic potentials. The coupling takes place at the isoB
plane position $Z_{rf}=\hbar \omega_{rf}/g_F\mu_B b' $. At this
position the energy separation between the different adiabatic
potentials is of the order of $ \hbar\Omega $ }}
\label{AdiabaticPotential1}
\end{figure}

The resulting adiabatic potentials are shown in a simple case on
figure \ref{AdiabaticPotential1}. The rf is applied in combination
with an inhomogeneous magnetic field $ B_z(0,0,z)=b'.z$ where $b'$
is a constant magnetic field gradient. The dressed states can be
decomposed into the bare states basis $|F=4,m_F=-4 ... +4\rangle$.
At the exact resonance position, the dressed states are linear
superposition of all the $|m_F\rangle$ substates, but far from the
interaction region, each dressed state is decomposed into a single
bare state \cite{Colombe04}. For this reason, we have labeled on
figure \ref{AdiabaticPotential1} (b) the different adiabatic
potentials branches with the name of the bare state they connect to
before and after the interaction region. An atom crossing the isoB
plane initially in the substate $|m_F\rangle$ will end in the
substate $|-m_F\rangle$. For example, atoms following adiabatically
the upper potential on figure \ref{AdiabaticPotential1}  (b) are
transferred from the initial $|m_F=-4\rangle$ bare state to the
final $|m_F=+4\rangle$ state. This spin flip has thus been made
possible by the absorption of 8 rf photons (as illustrated on figure
\ref{AdiabaticPotential1} (a) ) as atoms cross the resonance region.

Note that the adiabatic dressed states are eigenstates with energies
given by equation (\ref{equ1}) only for an atom at rest. Atoms cross
the interaction region with a velocity of $v\simeq 4$ m/s, which can
induce diabatic Landau-Zener like transitions between adiabatic
states. To reduce the probability of these transitions, the rf power
has to be high enough in order to increase the energy separation
between the different adiabatic potentials. The probability to
follow a single adiabatic potential can be described by a
Landau-Zener model and will be discussed in the last part of this
paper.

We have used this effect to realize a lens whose focal length can be
tuned by the rf frequency value : if atoms are prepared in the
$|^{7}S_{1/2},F=4,m_F= 4\rangle$ state, the first part of the lens
acts as a converging 2D lens. As they cross the isoB surface, atoms
are adiabatically transferred to the $|^{7}S_{1/2},F=4,m_F=
-4\rangle$ state and the second part of the lens acts as a diverging
2D lens. The combination of these two lenses realizes a doublet. By
changing the rf frequency value, the isoB plane position is changed
and the doublet focal length is varied. Atoms can also prepared in
the $|m_F=-4 \rangle$ sublevel, and an other doublet is realized.
\begin{figure}[h]
\centering
\includegraphics[width=3 in]{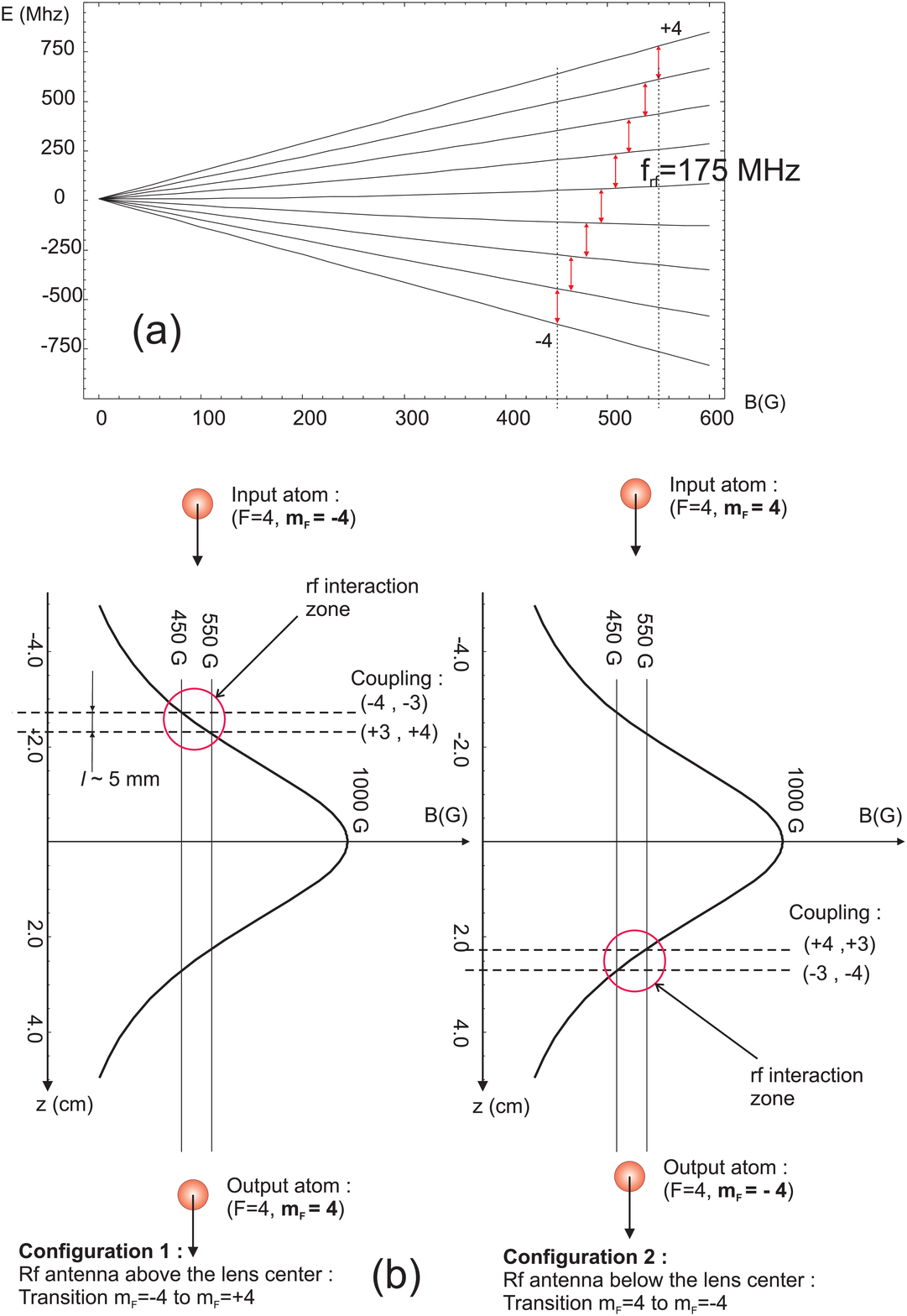}
\caption{\setlength{\baselineskip}{6pt} {\protect\scriptsize (a)
Zeeman diagram for the $ |F=4, m_F\rangle $ states. At a rf
frequency of 175 $MHz$ the coupling occurs at magnetic field values
between 450 $G$ and 550 $G$. (b) Two possible experimental
configurations, when the same rf frequency of 175 MHz is applied to
the magnetic lens. The magnetic field profile is plotted for 80 A in
the magnetic lens. Configuration 1 : The input atoms are polarized
in $|m_F=-4 \rangle$ state, and the rf antenna is located above the
lens center. Configuration 2 is exactly the opposite configuration.
The non linear Zeeman effect leads to a  spatial dispersion of the
isoB planes over 5 mm for these two experimental situations.}}
\label{NonLinearZeemanEffect}
\end{figure}

A difficulty arises as we want to shape magnetic potentials with
strong focusing properties, so that quite intense magnetic fields
are involved, and equivalently high rf frequency values. In that
case, the second order Zeeman effect becomes non negligible and
lifts the frequency degeneracy of the rf transitions between the
different $(|m_F\rangle,|m_{F}+1\rangle)$ pairs. For this
experimental study, we have used $\omega _{RF}/2 \pi$ ranging from
$100$ $MHz$ to $250$ $MHz$ corresponding to isoB field values
between $370$ $G$ and $655$ $G$. At such magnetic field intensities,
the second order Zeeman effect is already important and the coupling
between the successive magnetic sublevels is located at different
isoB plane, as shown on figure \ref{NonLinearZeemanEffect} (a). The
figure shows that for a rf frequency of 175 MHz, the resonance takes
place at a magnetic field value between 450 G for the
$|m_F=-4\rangle \leftrightarrow |m_F=-3\rangle $ transition and 550
G for the $|m_F=4\rangle \leftrightarrow |m_F=3\rangle $ transition.
As a consequence, the adiabatic transfer between the stretched
sublevels will be efficient only if atoms cross the different isoB
planes in the correct order. For an atom initially in the $|-4
\rangle$ state the transfer is effective if the atom crosses a
region of increasing field (positive gradient), for an atom
initially in the opposite state $|m_F=+4 \rangle $ the gradient has
to be negative. We have investigated the two possibilities leading
to two experimental configurations, as shown on figure
\ref{NonLinearZeemanEffect} (b). In the first one, atoms are
prepared in the $|m_F=-4 \rangle $ state, and the rf antenna is
located above the lens center. In the second one, atoms are prepared
in the opposite state and the rf antenna is located below the lens
center.

\begin{figure}[h]
\centering
\includegraphics[width=3 in]{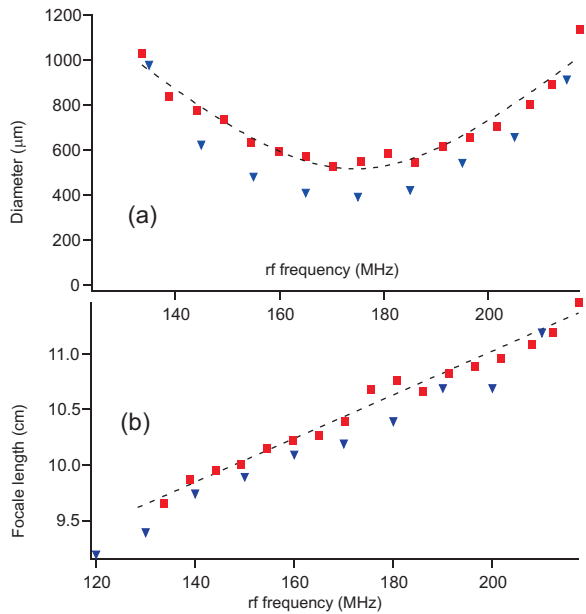}
\caption{\setlength{\baselineskip}{6pt} {\protect\scriptsize (a)
Variation of the cloud diameter at the probe position as a function
of the rf frequency. The current in the coil is 85 $A$. (b)
Corresponding variation of the focal length with the rf frequency.
The experimental data are deduced from (a) (see text). For (a) and
(b) : Experiment (squares) , numerical simulation (triangles). The
dashed lines are guides for the eye.}} \label{FrequencyDependance}
\end{figure}

\subsection{\label{DressedLens3}Experimental results}

We now turn to our experimental results. A first demonstration is
presented on figure \ref{FrequencyDependance}. The current amplitude
in the magnetic lens is set constant to 85 $A$. The antenna is
located above the lens center (configuration 1). Atoms are prepared
in the $ |F=+4, m_F=-4 \rangle $ state. The lens is first diverging,
then converging, as atoms are flipped from $|m_F=-4\rangle $ to
$|m_F=+4\rangle $ when they cross the rf interaction zone. The
resulting lens is a converging lens, but with a longer focal length
than without rf. We have measured the cloud diameter at the probe
position located 10.5 $cm$ below the lens center as a function of
the rf frequency. The results shown on figure
\ref{FrequencyDependance} (a) demonstrate that the magnetic lens
power can be tuned by changing only the rf frequency. The cloud is
focused at the probe altitude for a rf frequency value of 170 $MHz$.
The variation of the focal length with the rf frequency is presented
on figure \ref{FrequencyDependance} (b) : it has locally a linear
dependence with a slope of 125 $\mu m/MHz$. In this configuration,
the focal length value is always longer than the one obtained
without rf, equal to 7 $cm$ (in that case, atoms enter the lens in
$|m_F=+4\rangle$).  For figure \ref{FrequencyDependance}(b), the
experimental data have been deduced from figure
\ref{FrequencyDependance} (a) measurements using the following
observation: if the atom cloud is focused at a distance $f$ from the
lens center, with a minimal spot diameter $ D_{min}$, then the
diameter variation with the distance $\Delta z=(f-z)$ from the focal
point is given by $ D(\Delta z)=\sqrt{{D_{min}}^2+({D_{lens}}^2/f^2)
{\Delta z}^2}$ where $ D_{lens}$ is the cloud diameter at the lens
position. We have checked that all our experimental measurements
follow well this relation with the single parameter $D_{lens}=11$
$mm$ in correct agreement with a direct measurement (13 $mm $, see
figure \ref{Focalization} (a)). We present also on figure
\ref{FrequencyDependance} the results of the 3D numerical
simulations. The experimental data are well reproduced by numerical
simulations even if we have not introduced the non linear Zeeman
effect in our simulations. Indeed, as shown on figure
\ref{NonLinearZeemanEffect} (b), due to this effect, atoms have to
travel over $l=$ 5 $mm$ to be flipped from $|m_F=-4\rangle$ to
$|m_F=+4\rangle$. This distance remains nevertheless small compared
to the static field extension  (5 $cm$) and can be neglected in
numerical simulations. For these calculations, the rf effect is
taken into account by a sudden change of the force sign, as atoms
cross the isoB plane.

\begin{figure}[h]
\centering
\includegraphics[width=3 in]{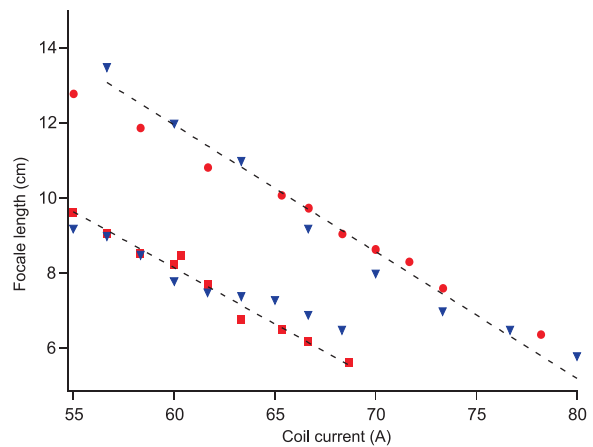}
\caption{\setlength{\baselineskip}{6pt} {\protect\scriptsize
Evidence for a rf-induced increase in the lens power. The focal
length is plotted versus the coil current for configuration 2. The
rf frequency is set to 210 $MHZ$. Experimental data (squares) = with
rf, (rounds)= no rf, and numerical calculations (triangles). The
dashed lines are guides for the eye. (The increase of the focusing
power of the lens with the rf is evident from a comparison of the
two sets of points.)}} \label{MoreConvergingLens}
\end{figure}

Configuration 1 is also interesting since it can be used to realize
a fast atomic shutter using no moving mechanical element. Atoms
initially prepared in the $ |m_F=-4 \rangle $ state are expelled
from the vertical axis when the rf-field is switched off. The atomic
signal at the probe beam position is then negligible. When the
rf-field is switched on, an intense signal appears. The switching
time is limited by the second order Zeeman effect, which is
responsible for a shift between the isoB planes coupling the
successive magnetic states (see figure \ref{NonLinearZeemanEffect}).
If we call $l$ the distance between the two extreme planes, the
switching time is equal to $\tau =l/v$ where $v$ is the mean atomic
beam velocity. For this experiment, with $ l\simeq5$ $mm$ (see
figure \ref{NonLinearZeemanEffect}(b)) and $v=4$ $m/s$ we got
$\tau\simeq1$ $ms$. To have a shorter switching time, an antenna
working at a smaller frequency could be placed in the lower field
region, where the second order Zeeman effect is much smaller. An
other possibility is to use atoms having no hyperfine structure and
therefore no non linear Zeeman effect, like $^{52} Cr$, which is
particularly well suited for atomic nanofabrication
\cite{Meschede2003}.

On figure \ref{MoreConvergingLens}, we show the results for
configuration 2: the rf antenna is located below the lens center and
atoms are initially prepared in the $ |m_F=+4 \rangle $ state. For
the experimental demonstration, the frequency was set to 210 $MHz$,
and we have measured the cloud diameter at the probe position,
located 8 $cm$ below the lens center, as a function of the current
in the magnetic lens. We have extracted from the data the effective
variation of the focal length as described for configuration 1. With
the rf ON, the cloud is focused 8 $cm$ below the lens center for a
current of 60 $A$. Without rf, the focal value jumps to 12 $cm$ for
the same current. Numerical calculations reproduce well the
experimental observations.

In configuration 2, the 'dressed lens' is always more focusing than
the 'bare lens', and the focal length decreases as the rf frequency
increases. As shown by further  simulations, the lens can be changed
from a converging lens to a converging mirror, by changing only the
rf frequency. For example, we have calculated numerically that for a
current of 100 A in the lens, and a rf frequency higher than 380
$MHz$, the dressed lens acts as a converging mirror. Such high rf
frequencies are beyond the values that we can reach with our rf
supply. This configuration is similar to the one used in
\cite{Bloch2001} where a resonator for atoms was demonstrated.

\subsection{\label{DressedLens4}Rf power requirements, Landau Zener
criterion}

Atoms cross the interaction zone with a velocity $v\simeq4$ $m/s$
giving rise to a non adiabatic probability transition. The
probability to have a single non adiabatic transition  between two
adjacent states $|m_F\rangle $ and $|m_F+1\rangle $ is given by the
Landau Zener formula \cite{Laundau32}:
\begin{equation}
     P_{LZ}=e^{-2\pi \Gamma(m_F,m_F+1)}
\end{equation}
 with
\begin{equation} \label{equ2}
 \Gamma(m_F,m_F+1) = \frac{\hbar \Omega^2(m_F,m_F+1)}{g_F \mu_B b' v}
\end{equation}
where $b'$ is the magnetic gradient and $\Omega(m_F,m_F+1)$ is the
Rabi frequency between $|m_F\rangle$ and $|m_F+1\rangle$. At low
field value, the coupling strength is given by \cite{Ketterle96}

\begin{equation} \label{equ3}
\Omega(m_F,m_F+1)=\frac{\mu_Bg_F}{4\hbar}B_{rf}\sqrt{F(F+1)-m_F(m_F+1)}
\label{toto2}
\end{equation}

We have checked that even at our magnetic field value (around 600
$G$) where the non linear Zeeman effect is non negligible, couplings
are not modified by more than 10 \% and we can use equation
(\ref{equ3}) to calculate them. On the other hand, the transitions
between the different $|m_F\rangle \leftrightarrow |m_F+1\rangle  $
are spatially separated and occur one after the other due to the
same effect. The atom transfer efficiency from $|-4\rangle$ to
$|+4\rangle$ is thus given by the product  of the probabilities of
the successive transitions between adjacent $|m_F\rangle$ substates
and is given by :

\begin{equation} \label{ProduitLandauZener}
P_{4\leftrightarrow(-4)}=\prod^{m_F=+3}_{m_F=-4}{(1-e^{-2\pi\Gamma(m_F,m_F+1)})}
\end{equation}

Using equation  (\ref{equ2}) and  (\ref{equ3}), we obtain that atoms
are adiabatically transferred if $h\Omega^2\gg\mu_Bg_Fvb'$. For
$v=4$ $m/s$ and $b'=200$ $G/cm$ we obtain that
$P_{4\leftrightarrow(-4)}>95 \%$ for $B_{rf}>200$ $mG$.

We turn now to the exact calibration of the rf magnetic field
amplitude $ B_{rf}$. We have verified that our single turn antenna,
in series with a 50 $ \Omega$ resistance, is well adapted to the rf
supply output impedance: the measured rf reflected power is less
than 10\%, showing that the antenna inductance gives a negligible
contribution to the total circuit impedance. We can then calculate
the AC current amplitude in the rf antenna for a given power, and
the magnetic field $B_{rf}$. At a power of 10 $W$, we estimate that
$B_{rf}=175$ $mG$ at the cloud center (1 $cm$ away from the rf coil
center). Note that the adaptation procedure works only if the rf
antenna diameter is small, otherwise the inductance increases and
the rf power is reflected by the antenna, so that there is almost no
current in the antenna. Our rf antenna has a diameter of 2 $cm$. Due
to its rather small size compared to the magnetic lens coil size,
the antenna position has to be adjusted to the mean rf frequency
value, in order to match its altitude with the isoB plane position
and thus to maximize the coupling strength. In fact, the magnetic
field  amplitude $B_{rf}$ is not well defined due to the transverse
extension of the cloud. Indeed, at the entrance of the magnetic
lens, the cloud radius is of the order of 4 mm (radius of a sphere
containing 80 $\%$ of the atoms). For a rf antenna located at 1 $cm$
from the cloud centre, the rf magnetic field amplitude varies
between 312 $mG$ and 97 $mG$ over the cloud extension, and is equal
to 175 $mG$ at the cloud centre as found before. This central value
gives a good estimation of the mean magnetic field seen by the
atoms, and can be used for our qualitative discussion on power
requirements.

\begin{figure}[h]
\centering
\includegraphics[width=3 in]{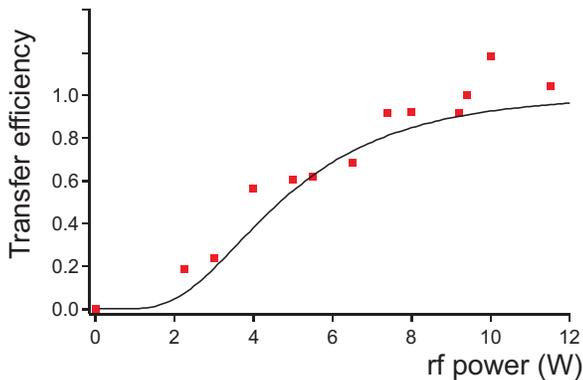}
\caption{\setlength{\baselineskip}{6pt} {\protect\scriptsize
Transfer efficiency as a function of the rf power. The rf frequency
was fixed to 130 $MHz$, the current in the lens to 95 A. Atoms are
initially polarized in $|m_F=-4\rangle$ state (configuration 1) and
we measure the number of atoms focused at the probe position and
normalized it (see text). The black line is obtained from
\ref{ProduitLandauZener}, the scaling between $\Omega^2$ and the rf
power being given by the experimental calibration.}}
\label{LandauZener}
\end{figure}

We have measured the transfer probability as a function of the rf
power (figure \ref{LandauZener}). Atoms are initially polarized in
the $|m_F=-4\rangle$ state. The probability is evaluated by
measuring the number of atom focused at the probe position for a
given rf power, normalized by the number of atom focused at the
probe position, with no rf, when atoms are initially polarized in
$|m_F=+4\rangle$ state. For a power of 10 W the transfer is close to
100\%. The data are in good agreement with the theoretical estimate
given by equation (\ref{ProduitLandauZener}).

\section{\label{DressedLens4}Conclusion} In conclusion, we have
demonstrated a new type of atomic lens by combining a permanent
magnetic lens with a rf field. The focal length of this atomic lens
can be finely tuned with the rf frequency value and the lens can be
changed from a less converging to a more converging lens, and even
could be changed to a converging mirror. The combination of rf
fields with static inhomogeneous magnetic fields gives new
properties to traditional atom optics elements like lenses, mirrors
and magnetic guides. The use of frequency combs or of rf sweeps can
add even more possibilities to generate more complex potentials
\cite{Courteille06}. We think that this innovative procedure can
join the atom optics toolbox since further improvements along this
scheme should result in the development of high quality coherence
preserving atomic adaptative lenses and mirrors. Futhermore the rf
dressing procedure can be readily combined with the well developed
integrated atom chip technology, to add coherent controls to
magnetic atoms chips \cite{zimmermann07}.




Acknowledgments: LPL is Unit\'e Mixte (UMR 7538) of CNRS and of
Universit\'e Paris Nord. We acknowledge financial support from
Conseil R\'egional d'Ile-de-France (Contrat Sesame), Ministère de
l'Enseignement Supérieur et de la Recherche, European Union
(Feder-Objectif 2), and IFRAF (Institut Francilien de Recherche sur
les Atomes Froids - MOCA project). We thank J.M. Fournier for many
fruitful discussions.








\end{document}